\documentclass[graybox,natbib,nosecnum]{svmult}
\bibpunct{(}{)}{;}{a}{}{,} 

\pdfoutput=1   

\usepackage{mathptmx}       
\usepackage{helvet}         
\usepackage{courier}        
\usepackage{type1cm}        

\usepackage{makeidx}         
\usepackage{graphicx}        
\usepackage{multicol}        
\usepackage[bottom]{footmisc}
\usepackage[normalem]{ulem}	
\usepackage{hyperref}  

\usepackage{amsmath}	
\usepackage{amssymb}
\usepackage{subfigure}

\usepackage{soul}   

\newcommand*\aap{A\&A}

\newcommand*\aj{AJ}

\newcommand*\apj{ApJ}
\newcommand*\apjl{ApJ}

\newcommand*\apjs{ApJS}

\newcommand*\icarus{Icarus}

\newcommand*\mnras{MNRAS}

\newcommand*\nat{Nature}

\newcommand*\earth{\oplus}

\makeindex             

\begin{document}

\title*{Formation of Super-Earths}
\author{Hilke E. Schlichting}
\institute{Hilke E. Schlichting \at University of California, Los Angeles, 595 Charles E. Young Drive East, Los Angeles, CA 90095, USA \& Massachusetts Institute of Technology, 77 Massachusetts Avenue, Cambridge, MA 02139, USA, \email{hilke@ucla.edu}}
%
%
\maketitle

\abstract{Super-Earths are the most abundant planets known to date and are characterized by having sizes between that of Earth and Neptune, typical orbital periods of less than 100 days and gaseous envelopes that are often massive enough to significantly contribute to the planet's overall radius. Furthermore, super-Earths regularly appear in tightly-packed multiple-planet systems, but resonant configurations in such systems are rare. This chapters summarizes current super-Earth formation theories. It starts from the formation of rocky cores and subsequent accretion of gaseous envelopes. We follow the thermal evolution of newly formed super-Earths and discuss their atmospheric mass loss due to disk dispersal, photoevaporation, core-cooling and collisions. We conclude with a comparison of observations and theoretical predictions, highlighting that even super-Earths that appear as barren rocky cores today likely formed with primordial hydrogen and helium envelopes and discuss some paths forward for the future.}

\section{Introduction}
Observations by the {\it Kepler} Space Telescope have led to the discovery of more than 4000 exoplanet candidates \citep{BK10,BR13}. These results provide us, for the first time, with a robust estimate of the relative abundances of different-sized planets with orbital periods of less than 100 days. The majority of the newly discovered planets reside well inside the orbit of Mercury around their respective host stars and have sizes between that of Earth and Neptune. Due to their sizes, this new class of planets is collectively referred to as super-Earths. It has already been established that super-Earths are ubiquitous and that about 50\% of all Sun-like stars harbor an exoplanet smaller than Neptune with orbital periods shorter than 100 days \citep{H12,F13}. This new class of planets is radically different from the planets in our own solar system, raising interesting questions concerning their nature and formation.

\section{Super-Earth Observations}

\subsection{Mass, Radius, \& Composition}
\begin{figure}
\centerline
\subfigure{\includegraphics[width=0.52\textwidth]{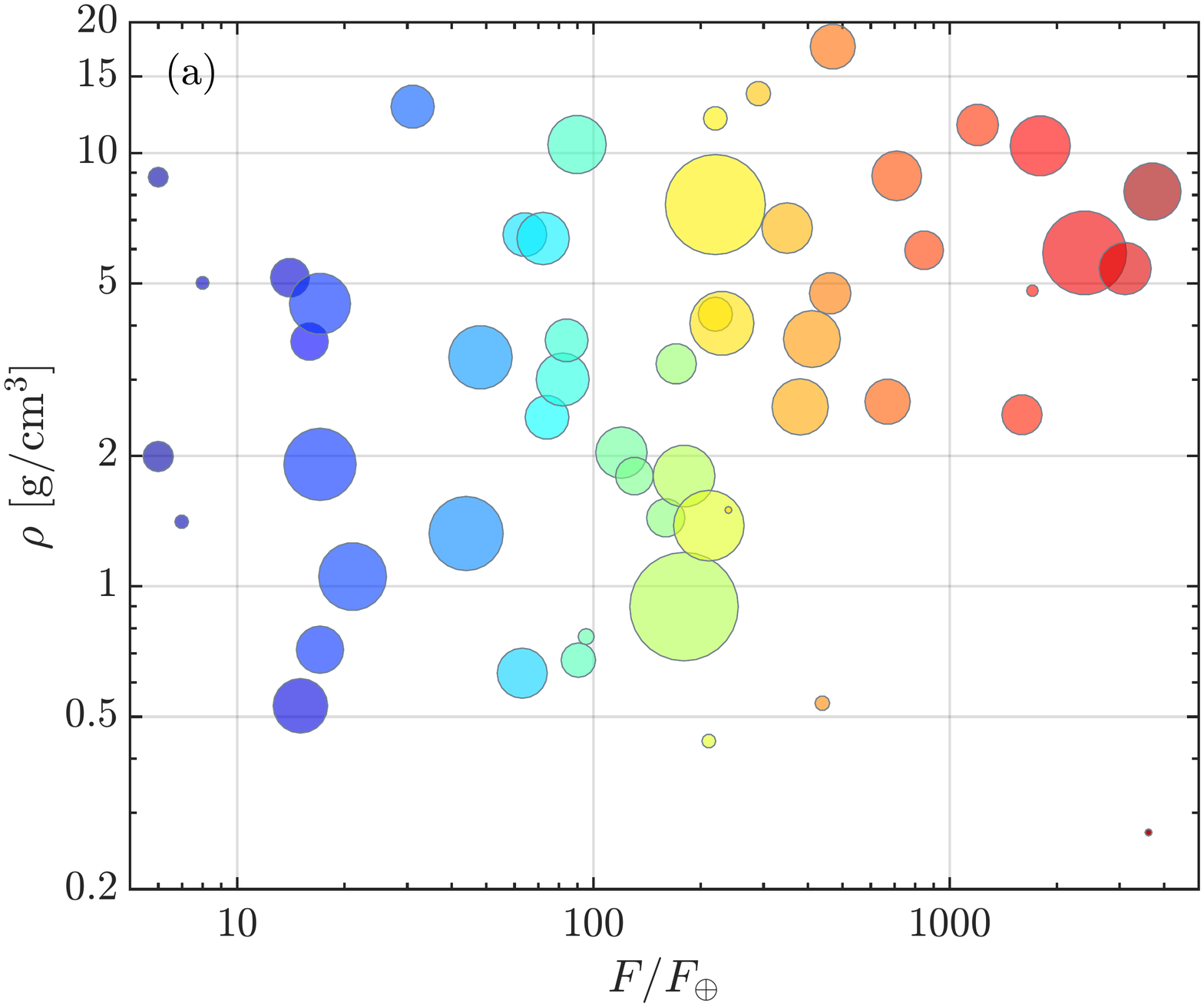}\hfil
\includegraphics[width=0.52\textwidth]{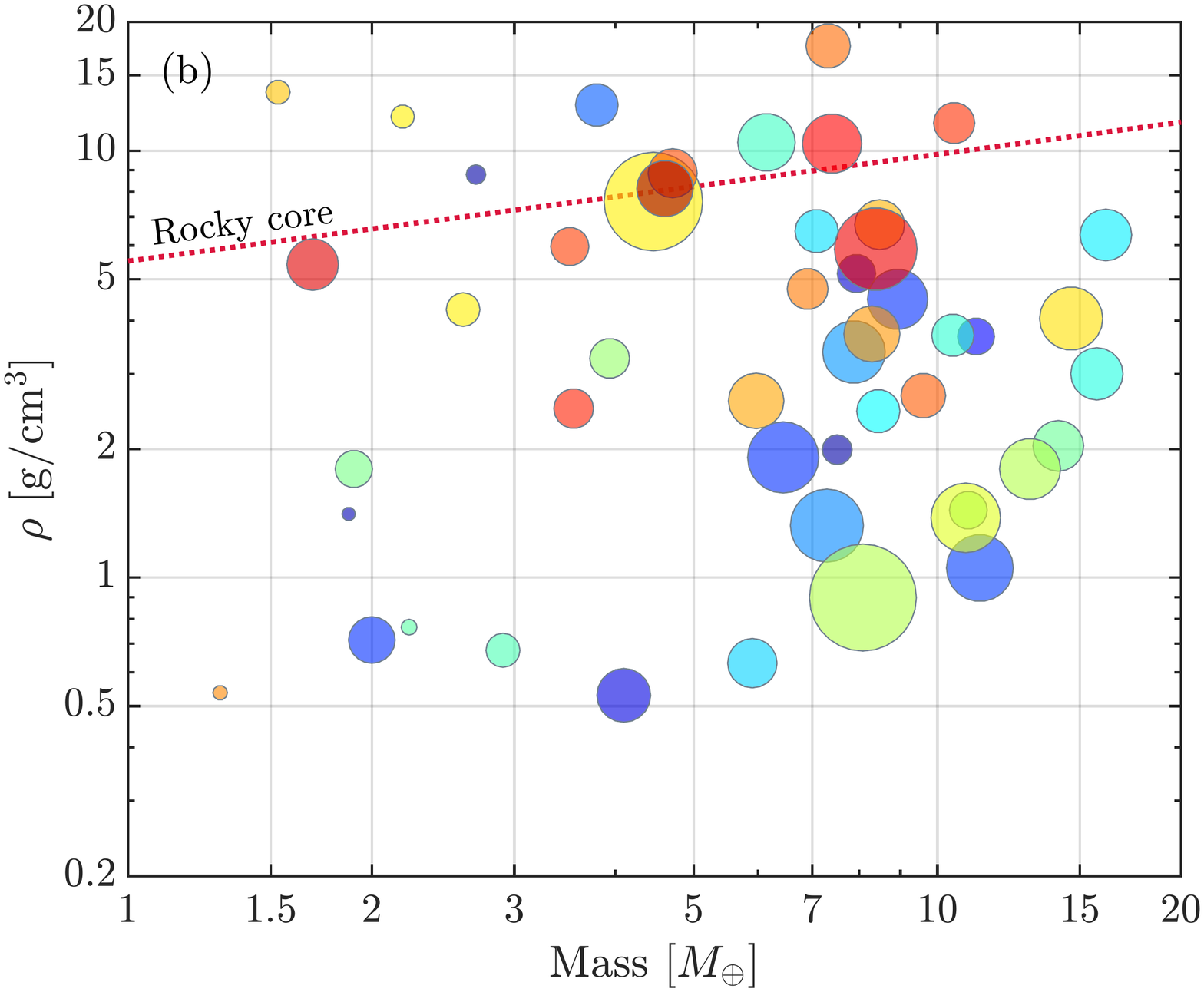}}\\
\subfigure{\includegraphics[width=0.50\textwidth]{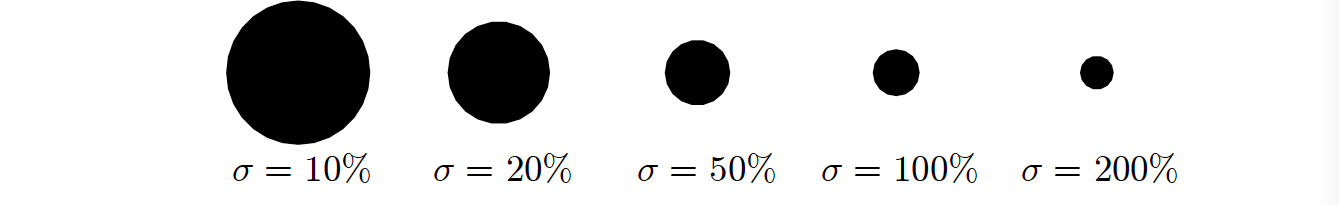}\hfil
\includegraphics[width=0.50\textwidth]{error4.png}
}
\caption{Diversity of bulk densities of detected exoplanets with radii less than 4 Earth radii (i.e., $R<4 R_{\earth}$). The surface area of each data point is inversely proportional to the $1 \sigma$ error of the density estimate, such that the most secure density measurements correspond to the largest points. The normalization of the error bars is shown at the bottom of the figure. The colors of the points represent the amount of flux received from the host star with the actual flux values shown in panel a). Panel a) displays the mean density as a function of flux, $F$, in units of the Earth flux, $F_{\earth}$. Panel b) shows exoplanet densities as a function of planet mass in units of Earth masses, $M_{\earth}$. Data are taken from \citet{WM14} and references therein. For reference, a mean density curve assuming a purely rocky planet \citep{SK07} is shown with a dotted red line. Figure from \citet{IS16}.}
\label{fig1}
\end{figure}

Thanks to {\it Kepler}, most super-Earths known to date were discovered by the transit method, which yields planet radii with uncertainties typically smaller than 10\%. Of the more than 4000 known super-Earths, about a hundred have reasonably well determined masses. The typical uncertainty in super-Earth masses is usually much larger than that of their radii. Super-earth masses have been predominantly determined by transit timing variations (TTVs) \citep{WL13,SFA13} (see also chapter on TTVs by Agol \& Fabrycky) and radial velocity measurements \citep[e.g.][]{WI14}. In both cases, super-Earth masses have, almost exclusively, been measured for planets residing in multiple-planet systems This is because the TTV method requires companions that lead to detectable gravitational perturbations in the orbital motions and because multiple-planet systems potentially offer more clues about their formation than their single-planet counter-parts and hence more resources were dedicated to studying them. For the sub-set of super-Earth systems for which both masses and radii are known, bulk densities can be calculated (see Figure \ref{fig1}) and inferences about their composition have been drawn \citep[e.g.][]{LF13,WM14,R15} (see also chapter on super-Earth composition by Rogers). However, in many cases, super-Earth radii are sufficiently large that even in the absence of any additional information the existence of an H/He envelope that contains a few percent of the planet's total mass can be inferred from the radius alone \citep[e.g.][]{LF14,WL15}. Figure \ref{fig1} shows the remarkably diverse bulk densities of super-Earths implying that super-Earths must have range of compositions even if they have similar masses. This is very different from any of the planets known in the solar systems, where planets of comparable mass (e.g. Uranus and Neptune, or Earth and Venus) share similar bulk densities and compositions.

\subsection{Orbital Architecture}
\begin{figure}
\centering
\includegraphics[width=90mm,angle=-90]{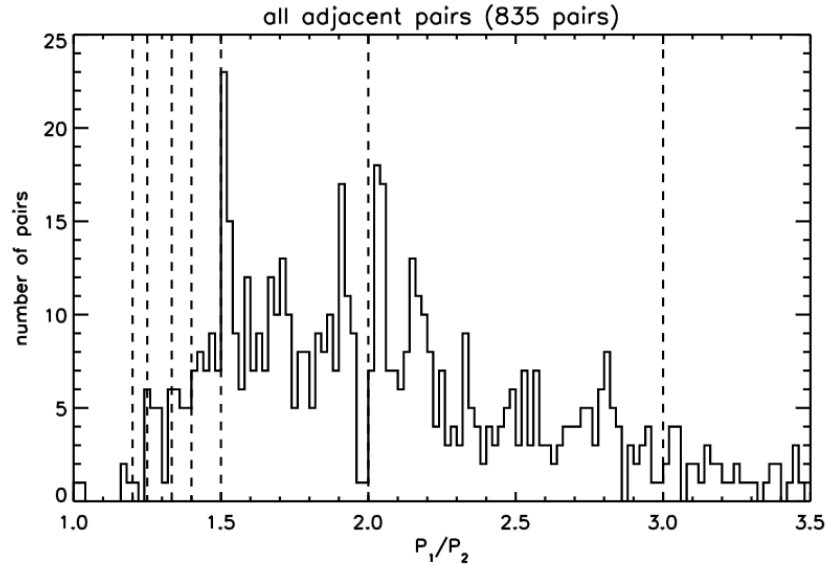}
\caption{Histogram displaying the orbital period ratio of all neighboring planet pairs recorded in the NASA Exoplanet Archive as of 2016 May 25. While there are some peaks in the histogram just outside several mean motion resonances (marked as dashed lines), most pairs seem to be completely unassociated with such resonances. Figure from \citet{PS17}.}
\label{Fig01}
\end{figure}

In contrast to hot Jupiters that rarely have a comparable sized companion, super-Earths frequently occur in multiple-planet systems \citep[e.g.][]{FL14}. As of July 2017, more than 580 multiple-planet systems have been discovered containing about 1500 planets, most of them super-Earths. These multiple-planet systems are usually tightly packed, with spacing ranging from 10-30 Hill radii \citep[e.g.][]{WM17}. Although some of these multiple-planet systems occupy mean-motion resonances \citep{MF16} most of them have period ratios unassociated with any resonance \citep{FL14} (see Figure \ref{Fig01}). This result seems surprising since gas giants, if they have companions, typically occupy mean-motion resonances \citep[e.g.][]{MB01} and because many super-Earths have significant gaseous envelopes implying that they formed in the presence of the gas disk and should therefore have experienced disk migration (see chapter by Nelson on `Planetary migration in protoplanetary discs') and efficient resonance capture. Several solutions have been proposed to this conundrum. 

Broadly speaking they fall into two categories: In the first, resonance capture is prevented due to turbulence in the disk \citep{R12} or by large eccentricities of the migrating planets \citep{B15}, which could be due to the mutual gravitational stirring  by the super-Earths among each other \citep{PS17}. In the second category of solutions, planets are efficiently captured into resonance but they escape on timescales that are shorter than the migration timescale between neighboring resonances due to overstable librations \citep{GS14,DB15} or the resonance chains are broken after the disk dispersal phase due to mutual gravitational stirring of the planets, leading to one to two collisions before establishing long term dynamical stability \citep{IS16,IO17}.

\section{Super-Earth Formation}
Having briefly summarized recent observations of the super-Earth population as a whole, we now turn to our attention to their formation scenarios. Broadly speaking, we can divide the formation process of super-Earths into three categories: (1) The formation of their planetary cores, (2) the accretion of their gaseous envelopes, and (3) atmospheric loss during and after formation. We discuss these three stages in detail below.

\subsection{1. Forming planetary cores}
In the absence of any migration or radial drift in the disk, the largest mass a planet or protoplanet can grow to by the accretion of solids is given by its isolation mass, which can be expressed as
 \begin{equation}
M_{iso}=\frac{(10\pi \Sigma a^2)^{3/2}}{3 M_{sun}^{1/2}},
 \end{equation} 
where $\Sigma$ is the disk mass surface density in solids, $a$ the semi-major axis and $M_{sun}$ the mass of the sun \citep{GL90,AR13}. This isolation mass is simply the sum of all material residing in the feeding zone of the planet. The feeding zone is the region over which a planetary embryo can directly accrete solids. It has a width of a few Hill radii, where the Hill radius, $R_H=a(M_{iso}/3M_{sun})^{1/3}$, is the distance from a body within which its own gravity dominates over the tides from the sun. The minimum-mass solar nebula (MMSN) \citep{H81} is the minimum mass that is needed to form the solar system and is constructed by taking the current planet masses, spreading them over annuli that extend half-way between the neighboring planets and enhancing the disk composition back to solar. Evaluating the isolation mass for the MMSN, yields $M_{iso} \sim M_{Mars}$ at 1.5 AU and roughly a Neptune mass at 20~AU. This implies that Mars, Uranus and Neptune may very well have formed as isolation masses, but that Earth must have undergone an additional stage of assembly consisting of mergers between dozens of roughly mars-sized planetary embryos (see chapter by Izidoro \& Raymond on the`Formation of Terrestrial Planets'). The maximum mass a planet can grow to when it is assembled with a phase of giant impacts is
\begin{equation}\label{eX2}
M_{max} \simeq \frac{[2^{5/2} \pi a^2 \Sigma (\rho/\rho_{sun})^{1/6}(a/R_{sun})^{1/2}]^{3/2}}{M_{sun}^{1/2}},
\end{equation}
where $\rho$ is the density of the planet and $\rho_{sun}$ the density of the sun \citep{S14}. Again, assuming a MMSN, we find from Equation (\ref{eX2}) $M_{max} \simeq M_{\earth}$. Equations (1) and (2) can be inverted and used to calculate the disk mass surface density in solids, $\Sigma$, needed to form the observed super-Earth population. \citet{S14} showed that if close-in planets formed in-situ as isolation masses, then standard gas-to-dust ratios yield corresponding gas disks that are gravitationally unstable for a significant fraction of systems, ruling out such a scenario. In addition, even with giant impacts \citep{RBM08,HM12,DLC16}, formation without migration requires disk surface densities in solids at semi-major axes of less than 0.1 AU with typical enhancements by at least a factor of 20 above the MMSN, which yields corresponding gas disks that are below, but not far from, the gravitational stability limit. In contrast, formation beyond a few AU is consistent with MMSN disk masses. This suggests that the migration of either solids or fully assembled planets is likely to have played a major role in the formation of close-in super-Earths and mini-Neptunes, or that local disk surface densities may not be representative of the whole disk and that planet formation may be confined to special localized regions in the disk. 

Interestingly, recent ALMA observations find disks displaying structure and rings. Furthermore, several works infer total gas disk masses close to the gravitational stability limit \citep[e.g.][]{vB17,PMS17}. Work by \citet{NK14} compared planet detection statistics with the measured solid reservoirs in T Tauri discs and concluded that planet formation is likely already underway at the few Myrs, with a large fraction of solids having been converted into large objects with low millimeter opacity and/or sequestered at small disc radii where they are difficult to detect at millimeter wavelengths.

\subsection{2. Gas-accretion}
\begin{figure}
\centering
\includegraphics[width=90mm,angle=0]{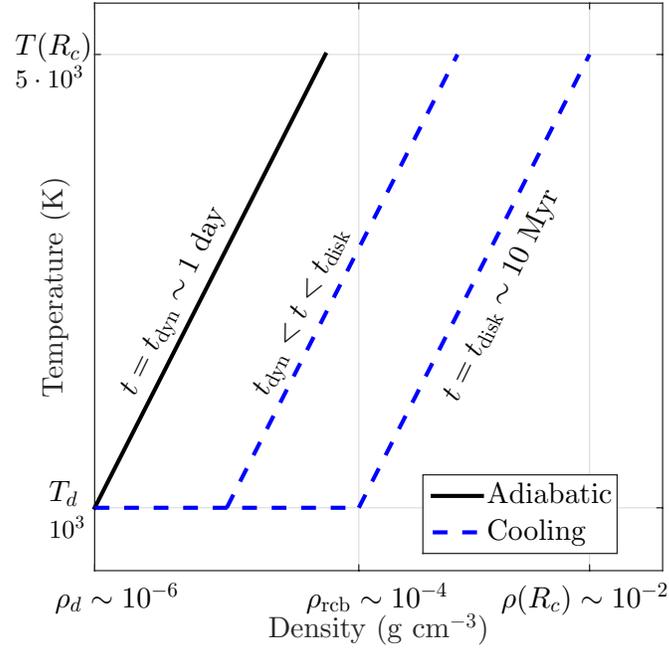}
\caption{Illustration of the evolution of the temperature and density profile of a super-Earth atmosphere during the nebular accretion phase. The initial adiabatic atmosphere (solid black line) is isentropic, while at later stages (2 successive profiles are plotted) the cooling (and accreting) envelope is characterized by a nearly isothermal, radiative outer layer, and a convective interior (dashed blue lines). Typical values of the density and temperature are given for reference. Figure after \citet{GSS16}.}
\label{fig12}
\end{figure}

The accretion of gas onto an already formed rocky core of mass, $M_C$, and radius, $R_C$, that is residing in a gas disk can be divided into two stages (see Figure \ref{fig12}). Initially, the gas adiabatically contracts onto the rocky core on dynamical timescales. In this case the density profile of the gas envelope is adiabatic and given by 
\begin{equation}\label{e1}
\frac{\rho(r)}{\rho_d}=\left(1+\frac{R_B'}{r}-\frac{R_B'}{r_{out}}\right)^{1/(\gamma-1)},
\end{equation}
where $\rho_d$ is the gas density in the disk, $\gamma$ the adiabatic index of the density profile, $R_B' \equiv (\gamma-1)(GM_C\mu)/(\gamma k_b T_{RCB})$, and $r_{out}$ the outer edge of the envelope, which is given by the smaller of the Bondi- or Hill Radius. Integrating Equation (\ref{e1}) yields typical envelope-to-core-mass fractions of 
\begin{equation}
f = \frac{M_{atm}}{M_{c}} \propto \rho_{d}
\end{equation}
and evaluates to about $f\sim 10^{-3}$ for minimum mass solar nebular type disks \citep{H81}. This is about an order of magnitude less than the typical super-Earth envelopes inferred from observations \citep[e.g.][]{WL15}.

Additional gas can be accreted by the core as its envelope starts to cool thereby lowering its entropy. As the envelope cools and contracts it develops an outer radiative region which connects to the convective interior envelope at the radiative-convective boundary, $R_{RCB}$ \citep{R06}. From then on the planet's accretion rate is governed by the cooling timescale of its gas envelope \citep[e.g.][]{LC15}. Specifically, the density profile of the gas envelope inside the $R_{RCB}$ is given by
\begin{equation}\label{eX1}
\frac{\rho(r)}{\rho_{RCB}}=\left(1+\frac{R_B'}{r}-\frac{R_B'}{R_{RCB}}\right)^{1/(\gamma-1)},
\end{equation}
where $\rho_{RCB}$ is the density at the radiative-convective boundary. This profile connects to an outer radiative, almost isothermal, envelope at the $R_{RCB}$. Essentially all the planet's mass is contained within $R_{RCB}$ because its density profile drops exponentially with a small scale height beyond it. The cooling time of the envelope is given by dividing the atmosphere's energy, $E$, by the internal luminosity $L$, (see e.g. \citet{GSS16} for details) where the energy of the envelope is given by 
\begin{equation}\label{e2}
E=-\frac{(\gamma-1)^2}{A(\gamma)\gamma(3-2\gamma)}\frac{GM_CM_{atm}}{R_C} \left(\frac{R_{RCB}}{R_C}\right)^{-(3\gamma-4)/(\gamma-1)}
\end{equation}
and the luminosity (cooling rate) of the envelope is
\begin{equation}\label{e3}
L=\frac{64 \pi}{3}\frac{\sigma T_{RCB}^4R_B'}{\kappa \rho_{RCB}},
\end{equation}
where $\sigma$ is the Stefan-Boltzmann constant and $\kappa$ the opacity at the radiative-convective boundary, which we take to be $\kappa \sim 0.1 \rm{cm}^2\rm{g}^{-1}$ \citep{AH01,FR08}. Combining Equations (\ref{e2}) and (\ref{e3}) yields the amount of gas that can be accreted as envelope by the time the disk disperses, which can, after some manipulation, be written as
\begin{equation}\label{e4}
f \simeq 0.02 \left(\frac{M_C}{M_{\earth}}\right)^{0.8} \left(\frac{T_d}{10^3 K}\right)^{-0.25} \left(\frac{t_{disk}}{1Myr} \right)^{0.5},
\end{equation}
where we made used of the fact that $T_d \simeq T_{RCB}$, since the outer region of the envelope is almost isothermal \citep{GSS16}. We note here that Equation (\ref{e4}) only logarithmically depends on the gas disk density, $\rho_d$, which is therefore omitted. This implies that in this regime the accretion rate is not regulated by the amount of gas present in the disk, but by the rate at which the gas can radiate away its gravitational energy and contract onto the core \citep{LC15}. This is in direct contrast to the initial adiabatic atmospheres for which $f$ scales linearly with the gas disk density. The gas-to-core-mass fraction given in Equation (\ref{e4}) should be regarded as an upper limit as any mechanism that heats the envelope can inhibit or diminish gas accretion. For example, giant impacts \citep{IS15}, the accretion of numerous planetesimals \citep{R06} and tides \citep{GS17} could all hamper gas accretion.

From Equation (\ref{e4}) we see that the gas-mass fractions in excess of $\sim 20\%$ are challenging to achieve at small separations from the host star and for typical disk lifetimes of a few Myrs. This suggests that most super-Earths did not turn into gas giants because the cores could not accrete enough gas to reach the runaway gas accretion stage during the disk lifetime and explains why super-Earths are much more abundant than (hot) Jupiters.

\subsection{3. Envelope loss during disk-dispersal, due to core cooling, by photoevaporation \& by collisions}
The envelope masses accreted during the gas disk phase, given by Equation (\ref{e4}), should not be regarded as final as several processes will act to reduce the envelope fractions during and after disk dispersal. Four such processed are discussed below in their likely order of occurrence in a super-Earth's life:

\subsubsection{3.1 Mass loss during disk dispersal}
As the gas disk dissipates, the density and pressure support around the planet decreases to zero. If the disk dissipation process, $t_{evap}$, is faster than the cooling time scale of the envelope, $t_{cool}$, the envelope will shed mass reducing the gas-to-core-mass fraction calculated in Equation (\ref{e4}). The Bondi radius sets the distance from a planet at which the thermal velocity of the gas exceeds the escape velocity from the planet. This implies that gas at the Bondi radius can flow into vacuum and escape from the planet. In order for the mass loss to continue, gas from further inside the envelope has to be supplied to the Bondi radius to replenish the escaping material. However, the temperature of the gas drops as it expands adiabatically. This implies that a constant supply of energy is needed to lift the gas out of the potential well such that it can reach the Bondi radius of the planet. The mass loss of the loosely bound outer layers of the atmosphere can be fueled by the heat escaping from the contracting inner layers of the atmosphere \citep{GSS16,OW16}. The ratio between the envelope contraction timescale and the atmospheric loss timescale is simply the ratio of the thermal energy available for cooling in the envelope (that is dominated by the inner layers) and the gravitational binding energy of the outer regions of the atmosphere and can be written as
\begin{equation}\label{e5}
\frac{t_{\rm evap}}{t_{\rm disk}}=\frac{t_{\rm evap}}{t_{\rm cool}}=\frac{E_{\rm evap}}{E_{\rm cool}}=\left(\frac{R_{\rm RCB}}{R_C}\right)^{-(3-2\gamma)/(\gamma-1)},
\end{equation}
\citep{GSS16}. From Equation (\ref{e5}) we see that atmospheric mass loss proceeds as long as $R_{RCB} \gg R_C$. This implies that super-Earths lose their outer envelopes which, typically contain, depending on the value of $\gamma$, 25\% to 70\% of the total gas mass. At the same time the radius of the envelope contracts until it become comparable to $R_C$. All this evolution happens rather rapidly on a timescale comparable to the disk life time, which is typically a few million years. In summary, as the gas disk disappears, the envelope rapidly loses dozens of percent in mass and shrinks to a radius comparable to the planet's core radius in a few million years. Equation (\ref{e4}) therefore gives only the expected gas-to-core-mass fraction due to accretion while the gas disk is present. The actual gas-to-core mass fraction at the end of disk dispersal is
\begin{equation}\label{e6}
f \simeq 0.01 \left(\frac{M_C}{M_{\earth}}\right)^{0.44} \left(\frac{T_d}{10^3 K}\right)^{0.25} \left(\frac{t_{disk}}{1Myr} \right)^{0.5},
\end{equation}
 \citep[e.g.][]{GSS16}.\\

\subsubsection{3.2 Core-powered mass loss}
Once the envelope has shed the outer layers it enters the `thin' regime where the thickness of the atmosphere is comparable to or less than the planet's core radius, $R_C$. The fate of the subsequent evolution of the envelope from now on is determined by the amount of mass and energy contained in the envelope compared to the core \citep[e.g.][]{IH12,GSS16}. Atmospheric mass loss will continue if sufficient energy can be released from the envelope and core to lift the gas out of the planet's potential well.
The energy that is available for cooling of the envelope in the thin regime is
\begin{equation}\label{e7}
E_{cool}=gR_{\rm RCB}\left(\frac{\gamma}{2\gamma-1}M_{\rm atm}+\frac{1}{\gamma}\frac{\gamma-1}{\gamma_c-1}\frac{\mu}{\mu_c}M_C\right),
\end{equation}
where $g\equiv GM_c/R_c^2$ is the surface gravity and $\mu_c$ and $\gamma_c$ are the core's molecular weight and adiabatic index, respectively. Equation (\ref{e7}) assumes that the core is molten and roughly isothermal, which is valid for super-Earths given their massive envelopes and their proximity to their host stars. From Equation (\ref{e7}) we can see that the energy available form cooling is dominated by the core for light atmospheres ($f \lesssim \mu/\mu_C$) and by the thermal and gravitational energy of the envelope for heavy atmospheres ($f \gtrsim \mu/\mu_C$) \citep{GSS16}. 

\begin{figure}
\centering
\includegraphics[width=120mm,angle=0]{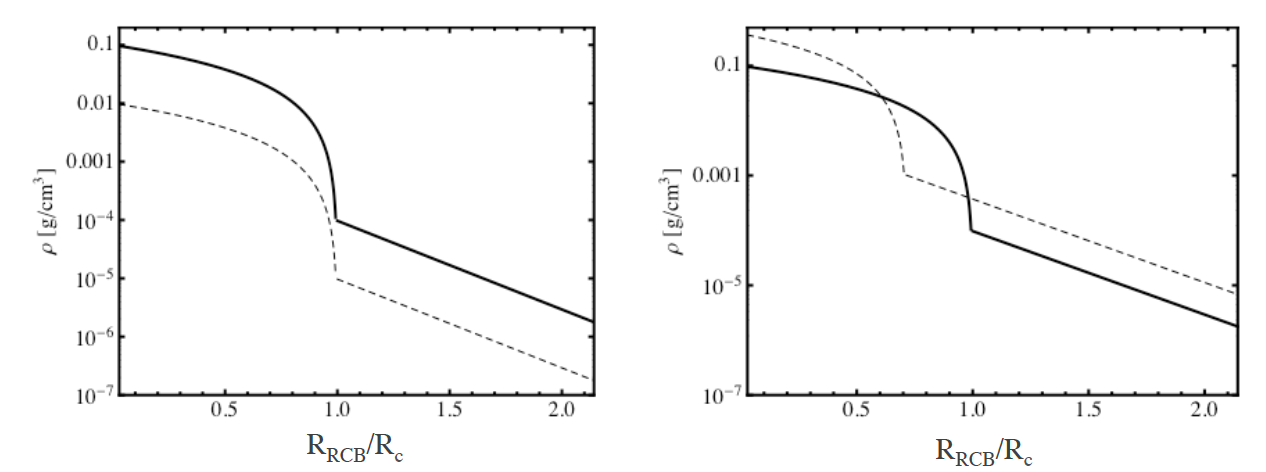}
\caption{Illustration of the evolution of the density profile of a super-Earth atmosphere after the gas disk has disappeared. The solid line corresponds to the time when radiative-convective boundary of the atmosphere, $R_{RCB}$, has shrunk to $R_{c}$, which happens on a timescale of about 10~Myrs due to rapid loss of the outer envelopes during the disk dispersal phase. The dashed line corresponds to a later time in the super-Earth evolution. The case for heavy atmospheres (i.e. $M_{atm}/M_{c} \gtrsim 5\%$) is shown on the right and the evolution of light atmospheres (i.e. $M_{atm}/M_{c} \lesssim 5\%$) is shown in the left panel. Heavy atmospheres cool and shrink once the $R_{RCB}$ contracts to about $R_{C}$; light atmospheres, on the other hand, experience core powered mass loss, which decreases the envelope mass over time while keeping $R_{RCB}$ constant at about $R_{c}$.}
\label{fig3x}
\end{figure}

For an Earth-like composition core and a hydrogen and helium dominated atmosphere, the transition between heavy and light envelopes occurs at envelope-to-core mass fractions of about 5\%. 
The two different evolution scenarios of light and heavy envelopes are illustrated in Figure 4. Since the gravitational binding energy of thin atmospheres is $E=GM_cM_{\rm atm}/R_C=gM_{\rm atm}R_C$, heavy atmospheres can cool and contract without any additional mass loss (see Figure \ref{fig3x}), whereas light-atmospheres can be lost completely because the thermal energy from the core exceeds the binding energy of the envelope. Because the energy available form cooling of light atmospheres is dominated by their core, the envelopes are not able to contract as the planets cools (see Figure \ref{fig3x}) and atmospheric mass loss continues with $R_{RCB} \sim R_{C}$. This implies that the energy required to lose the envelope decreases with time while the energy available for cooling from the rocky core remains almost constant. As a result, light atmospheres can be lost completely. 

There are two things can ultimately save light atmospheres: The first is the timescale for atmospheric loss which is set by the finite escape rate of molecules from the Bondi radius. This mass-loss timescale is given by
\begin{equation}\label{e8}
t \sim \frac{R_B'}{c_s} \left(\frac{R_{RCB}}{R_B'}\right)^{(3\gamma -4)/(\gamma-1)} \exp\left(\frac{R_B}{R_{RCB}} -1\right),
\end{equation}
and it can exceed the age of the system, such that light atmospheres can survive to the present day simply because their mass-loss timescales exceed several Gyrs \citep{GSS16,OW16}. Choosing an age of a few Gyrs, Equation (\ref{e8}) can be inverted to yield a condition on the mass and equilibrium temperature for a super-Earth to be able to keep it's light atmosphere. This condition can be written as
\begin{equation}\label{e9}
\frac{M_C}{M_{\earth}} \gtrsim 6.3 \left(\frac{T_{eq}}{10^3~\rm{K}}\right)^{4/3},
\end{equation}
\citep{GSS16}.

The second process that can save the light atmosphere is that the cores will ultimately cool to low enough temperatures that an insulating solid curst can form. Such a crust can prevent or even shut-off efficient heat transport from the core to the envelope and hence terminate the atmospheric loss. However, super-Earths and their envelopes are sufficiently massive that the formation of a solid curst can be ignored until late times (typically Gyrs) and may not happen at all in certain systems. In addition, heat generated by radiactive decay in core and by impacts can further delay the formation of a solid crust.

\subsubsection{3.3 Photoevaporation}
In addition to the atmospheric loss discussed above, planetary atmospheres will be eroded by photoevaporation \citep[e.g.][]{LF12,OJ12}. In contrast to envelope loss due to the core cooling of the planet, which is related to the stellar bolometric luminosity, since this sets the planet's equilibrium temperature, which in turn determines both its cooling and mass loss rates, photoevaporation is powered by the high-energy tail of the stellar radiation. Planetary envelopes are evaporated by ionizing photons that release energetic electrons and that in turn heat the atmosphere to temperatures above the escape velocity. As a result, the hot gas escapes the planet's potential well, provided that the cooling of the gas is slow enough. As discussed before, in order for the mass loss to continue, gas from further inside the envelope has to be supplied to the Bondi radius to resupply the escaping material. A constant supply of energy is needed to lift the gas out of the potential well because the temperature of the gas drops as it expands adiabatically. This is generally referred to as energy-limited escape. The widely used energy-limited model for photoevaporation linearly connects the high-energy flux to the gravitational energy of the escaping material  \citep[e.g.][]{LF12,OJ12}. The photoevaporation flux is commonly parameterized as $L=4\pi R_{RCB}^2 \sigma T^4_{RCB} \eta$, where $\eta \sim 10^{-4}$ accounts for both the evaporation efficiency, which is typically assumed to be $\sim 0.1$, and the fraction of ionizing radiation of the total bolometric stellar flux, which is assumed to be constant at $\sim 10^{-3}$ during the first 100~Myrs. This yields a photoevaporation timescale of
\begin{equation}\label{e10}
t_{evap} = \frac{M_{atm} g R_c}{4 \pi R_c^2 \sigma T_{RCB}^4 \eta}.
\end{equation}
From Equation (\ref{e10}) we see that $t_{evap} \propto M_{atm}$ implying that massive envelopes can survive photoevaporation, provided that such massive envelopes could have been accreted during the disk phase in the first place \citep[e.g.][]{LF12,IS15,LC15}. Accounting self-consistently for both the accretion of the envelope and the loss of the outer layers during disk dispersal yields the following condition for super-Earths to keep their massive envelopes
\begin{equation}\label{e11}
\frac{M_c}{M_{earth}} \gtrsim 7.7 \left( \frac{T_{eq}}{10^3 \rm{K}} \right)^{2.22}
\end{equation}
\citep{GSS16}. Equation (\ref{e11}) demonstrates that super-Earths need to be massive and/or have low equilibrium temperatures in order to keep their envelopes.

Finally, we note that the prescription used here for energy-limited escape by photoevapoation is, of course, an approximation and that, in addition, other mass loss regimes exist. For example, when the recombination timescale is short compared to the flow timescale, the loss is recombination-limited and when the planet's potential well is sufficiently shallow the mass loss can be photon-limited \citep{OA16}. Which of these mass loss regimes applies depends on the high-energy photon flux from the host star and the properties of the planet.

\subsubsection{3.4 Envelope loss due to late collisions}
Finally, in addition to envelope loss due to disk-dispersal, the planet's own core cooling history and photoevaporation, envelopes can be lost by giant impacts between super-Earths. Giant impacts that occur after the gas disk has dissipated may be common because super-Earths must have formed in the presence of gas disks and their dynamical interaction with the gas disk is expected to have resulted in migration and efficient eccentricity damping (see chapter by Nelson on `Planetary migration in protoplanetary discs'). This leads to densely-packed planetary systems. As the gas disk dissipates, mutual gravitational excitations between the planets cause their eccentricity to growth culminating in one or two giant impacts before reaching long-term orbital stability \citep{IO17}. Giant impacts maybe a particularly attractive solution for explaining tightly pack multiple planets systems with very similar core masses but vastly different envelope fractions and hence bulk densities \citep{L15,IS16,HC17}, like, for example, Kepler-36 \citep{CAC12}. Systems like Kepler-36 are challenging to explain as a result of gas accretion and evolution alone. This is because the envelope fractions accreted (see Equations (\ref{e4}) and Equations (\ref{e6})) are primarily functions of the core mass, equilibrium temperature and gas disk lifetime and we know that the core mass and equilibrium temperature are similar for both planets and that the gas disk lifetime should have been the same for planets in a given system. Striping atmospheres by impacts therefore presents an attractive solution to explaining tightly packed super-Earth systems with vastly different bulk densities (see Figure \ref{fig1}).

\begin{figure}
\centering
\includegraphics[width=90mm,angle=0]{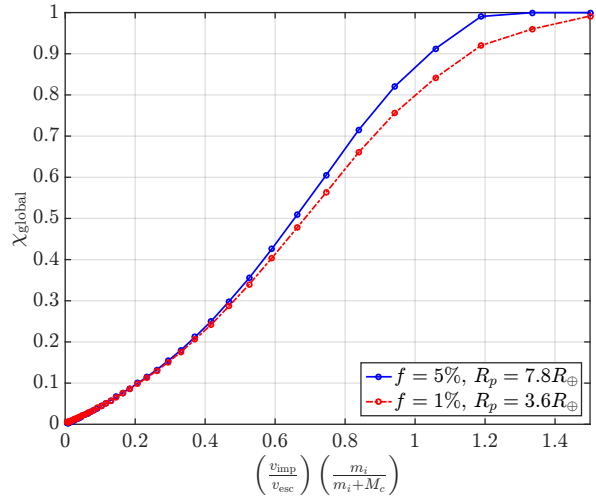}
\caption{Atmospheric mass loss fractions, $\chi_{global}$, as a result of giant impacts in super-Earth systems as a function of impactor momentum, $(v_{imp}/v_{esc})(m_i/(M+m_i)$. Shown are results for super-Earths with 5\% (blue) and 1\% (red) envelope fractions with core masses of $4M_{\earth}$ and a face-on impact geometry. A giant impact between similar sized planets can easily lead to loss of at least half of the total envelope, if not more. Figure from \citet{IS15}.}
\label{fig13}
\end{figure}

When a giant impact between two super-Earths occurs, it creates a strong shock at the impact site that propagates though the entire planet. This in turn leads to a global ground motion of the core which results in a second shock being launched into the planet's envelope. The part of the envelope that is accelerated by the shock above the escape velocity from the planet is lost (see \citet{IS15} for details). Figure \ref{fig13} shows the envelope mass loss fraction as a function of impactor momentum for head on impacts. Since impactors are accelerated to at least the escape velocity upon impact, a collision between comparable sized super-Earths will lead to loss of at least half of the total envelope \citep{IS16}. This in turn will modify the bulk density of super-Earths by a factor of a few when accounting for the subsequent thermal evolution over Gyr timescales \citep{IS16}.

Finally, the atmospheric mass loss shown in Figure \ref{fig13} is likely an underestimate of the true mass loss because even parts of the original envelope that was not immediately lost is, due to its large inflated post-collision radius, susceptible to subsequent loss by Parker winds and photoevaporation \citep{L15}. Furthermore, Figure \ref{fig13}, shows the atmospheric mass loss in head-on collisions, assuming that super-Earth systems experienced only one collision that result in a merger after the gas disk dissipated such that the system can reach long term dynamical stability. However, it is possible that the super-Earth systems have undergone several hit-and-run collisions, i.e. collisions that do not lead to a merger, before a final impact merger occurs. If that is the case, the atmospheric mass loss can be enhanced significantly, since hit-and-run collisions can lead to several rounds of atmospheric loss without reducing the number of planets in the system \citep{HC17}.

\section{Comparing observations \& theory: Implications for the origin and evolution of super-Earths}
Having given a brief outline of possible formation and evolution scenarios of super-Earth systems, we dedicate the last section of this chapter to a comparison of the observed super-Earth population and the results derived above.

\subsection{Global properties of super-Earth systems}
\begin{figure}
\centering
\includegraphics[width=0.7\textwidth]{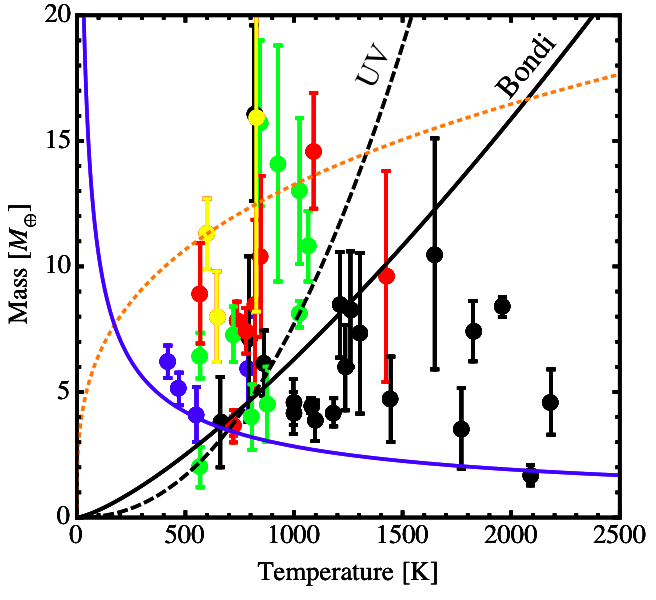}
\caption{Observed super-Earth population with known masses and radii from \citet{WM14}. The planets are color-coded according to their gas-mass fraction, $f$, with black corresponding to $f<0.3\%$, red to $0.3\%<f<1\%$, green to $1\%<f<5\%$, blue to $5\% < f < 10\%$ and yellow to $f>10\%$. The solid-blue line represents planetary core masses needed to accrete and retain envelope masses fraction of $f=5\%$ as a function of equilibrium temperature from the star and is given by Equation (10). Planets above the dotted-orange line are expected to undergo run-away gas accretion becoming Jupiters instead of super-Earths. The lines labeled `UV' and `Bondi' correspond to limits for complete atmospheric loss due to photo-evaporation (see Equation (\ref{e11})) and loss due to thermal energy from the cooling core and envelope (see Equation (\ref{e9})), respectively. Figure updated from \citet{GSS16}.}
\label{fig2}
\end{figure}

Since gas accretion, the thermal evolution of the envelope and its atmospheric mass loss all depend on the super-Earth's mass  and the equilibrium temperature (see Equations (\ref{e6}), (\ref{e9}) and (\ref{e11})), is instructive to summarize and compare all these results in a single plot that displays the envelope fraction as both a function of super-Earth mass, which is to first order the same as the planet's core mass, and equilibrium temperature (see Figure (\ref{fig2})). Figure \ref{fig2} displays the observed super-Earth population with known masses and radii from \citet{WM14}. The planets are color-coded according to their gas-mass fraction, $f$, with black corresponding to $f<0.3\%$, red to $0.3\%<f<1\%$, green to $1\%<f<5\%$, blue to $5\% < f < 10\%$ and yellow to $f>10\%$. The solid-blue line shows the relation for gas accretion as a function of core mass and equilibrium temperature as given by Equation (\ref{e4}) at the end of disk dispersal and is plotted for $f=5\%$. The black-dashed line corresponds to the relation for complete envelope loss by photoevaporation given in Equation (\ref{e11}), and the solid-black line the corresponding relation for complete envelope loss by core cooling over a timescale of Gyrs (see Equation (\ref{e9})). The region below the black lines should be devoid of super-Earths with significant gaseous envelopes because planets in this parameters space could not have retained their primordial envelope, which is indeed consistent with observations. This implies that the concentration of super-Earths below the black lines with no significant gaseous envelopes should not be interpreted as having formed as rocky cores without any H/He envelopes, but instead suggests that they may be part of the super-Earth population that formed in the presence of a gas disk and with primordial H/He envelopes and that they were stripped of their primordial atmospheres. 

On the other hand, super-Earths located above the black lines are expected to have retained their primordial atmospheres, which is also consistent with observations. However, Equation (\ref{e6}), predicts, for a given gas disk lifetime, a segregation in the core mass - equilibrium temperature space for planets with different envelop fractions, but this is not born out by the data. The cause for this may be threefold. First off all, there is significant uncertainly in the planet's masses, as indicated by the one sigma error bars shown in Figure \ref{fig2}, so any trends with planet mass and envelope fraction will be strongly blurred due to the large errors in super-Earth masses. Second, as discussed above, the envelope fractions accreted are a function of disk lifetime which can vary from system to system. Thirdly, giant impacts that occur after the gas disk disperses will erase the trends predicted from accretion for all planets that were part of a collision, since giant impacts cause significant atmospheric loss. Furthermore, since giant impacts result in net atmospheric loss, they may explain why so many of the green ($1\%<f<5\%$) and red ($0.3\%<f<1\%$) points lie above the blue line corresponding to 5\% envelope fractions, rather than below, suggesting that they could correspond to super-Earths that had more massive envelopes that were subsequently diminished by giant impacts.

\begin{figure}
\centering
\includegraphics[width=110mm,angle=0]{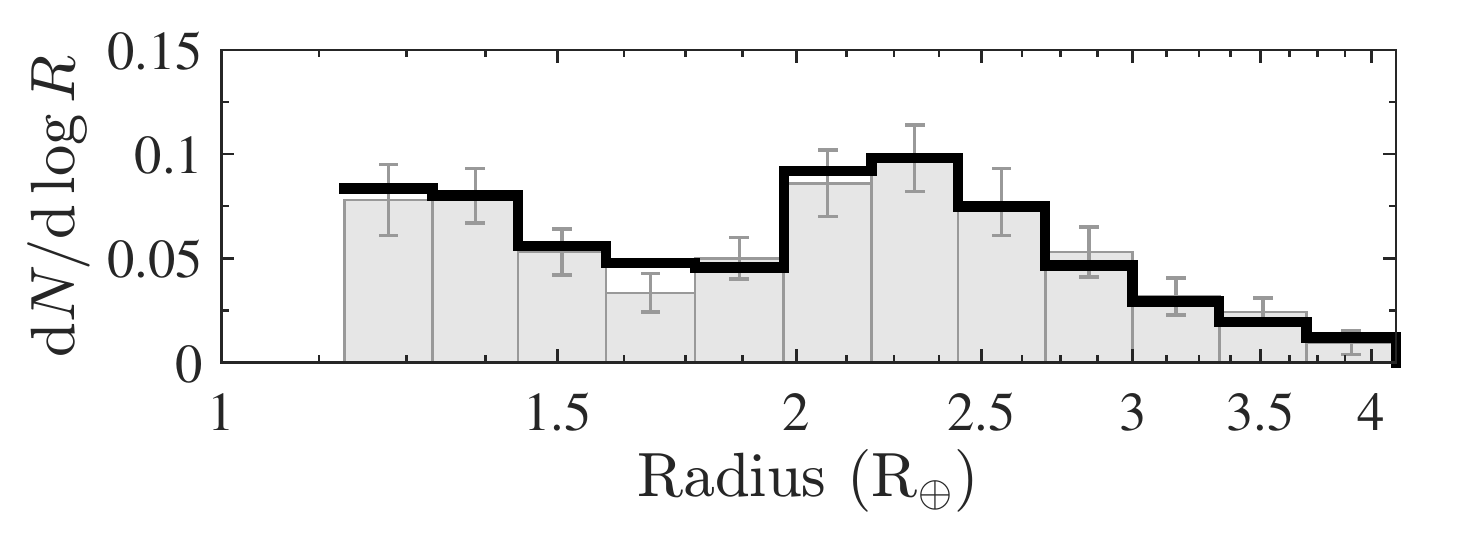}
\caption{Exoplanet radius distribution. The grey histogram corresponds to the observed distribution from \citep{F17}. The black solid line corresponds exoplanet radii after 3 Gyrs of evolution accounting for the core-cooling mass loss described in section 3.2 above. The initial planet mass distribution was chosen to be consistent with that reported by \citet{WI14} and the distribution of initial gas envelopes was chosen such that $M_{atm}/M_{core} \propto M_{core}^{1/2}$ as predicted from planet formation models (see Equation (\ref{e6})). Figure after \citet{GSS17}.}
\label{fig14}
\end{figure}

A second, somewhat complementary, way to examine the super-Earth population is to examine their radius distribution shown in Figure \ref{fig14}. The grey histogram corresponds to the observed distribution from \citep{F17} with Poisson error bars. The data demonstrates, as it had been previously noticed by \citet{OW13}, that there is a deficit of intermediate sized super-Earth with radii of about $1.5 - 2.0 \rm{R_{\earth}}$. The reason for this deficit has been attributed to atmospheric erosion by photoevaporation \citep{LF13,OW13,JM14,CR16,OLC17}, i.e. the line that is labeled as `UV' in Figure \ref{fig2}.  However, the paucity of planets in the  $1.5 - 2.0 \rm{R_{\earth}}$ can be equally well explained by atmospheric loss driven by cooling of the planet's core  \citep[e.g][]{IH12,GSS16}. In fact the solid-green line in Figure \ref{fig14} corresponds to the radius distribution of planets that only undergo thermal evolution and mass loss due to core cooling over Gyr timescales \citep{GSS17}. Since the luminosity of the cooling core can erode light atmospheres while preserving heavy ones, it produces a deficit of intermediate size planets. 

It is challenging to disentangle the importance of these two mass loss mechanisms observationally, since they have a similar dependence on equilibrium temperature and planet mass (see Figure \ref{fig2} and expressions given in Equations (\ref{e9}) and (\ref{e11})). The only region (enclosed by the two black lines in Figure \ref{fig2}) where these two loss mechanism differ visibly is for massive-high temperature planets. Nonetheless, observations of planets orbiting different stellar types may be able to distinguish between envelope loss dominated by photoevaporation, which is powered by the high-energy tail of the stellar radiation over the first 100~Myrs, and core cooling driven mass loss, which is dictated by the bolometric flux of the host star. Since there is a large scatter in the high-energy flux of stars with the same mass \citep{TJ15}, photoevaporation should lead to a less distinct desert in the radius distribution of short-period super-Earths than mass-loss driven by core cooling.

\subsection{Discussion \& Conclusions}
Since the discovery of the first super-Earths, considerable progress has been made illuminating their formation histories and origins. In contrast to the terrestrial planets in our solar system, super-Earths must have formed in the presence of a gas disk in order to accrete the H/He envelopes that are inferred for a significant faction of them. The planet's gravitational interaction with the gas disk likely resulted in migration implying that super-Earths, or at least their building blocks, probably did not form where we observe them today. However, whether this migration covered several  AU or only a factor of a few of their current locations remains to be established. It has been suggested that the planet population outside 1 AU may have been responsible for dynamical shaping the close-in super-Earth systems that we see today \citep{IRM15}. Characterizing the longer period exoplanet population residing outside the super-Earth population is certainly an exciting prospect for illuminating their formation. In addition, spectroscopic measurements of super-Earth atmospheres will shed light on their composition in the near future and may provide valuable clues to their formation locations as well. 

As summarized in this chapter, progress has been made in understanding the gas accretion phase onto super-Earth cores and the various atmospheric mass loss mechanisms at work during and after the dispersal of the gas disk. Our current theoretical understanding is consistent with observations and suggests that even super-Earths that appear purely as rocky cores today may have originally formed with primordial H/He envelopes. In addition, the large spread in super-Earth envelope fractions for almost identical core masses suggests that giant impacts after the gas disk dissipated may have played a role in shaping super-Earth systems. Furthermore, since giant impacts result in net atmospheric loss, they may also explain why many super-Earths are observed with envelope fractions below those predicted by accretion  and evaporation alone, suggesting that these planets too could have had more massive envelopes that were subsequently diminished by giant impacts. Despite these advances, several interesting questions remain. The gas accretion models discussed here assume spherical symmetry and it has been suggested that the multi- dimensional gas flow around an accreting super-Earth may alter the one-dimensional results \citep{DAB13,F15,O15a,O15b}. However, no agreement about the magnitude of this effect has been reached and further work investigating these issues is ongoing.  

Finally, super-Earths with sufficiently large radii that required a 5-10\% mass fraction in H/He envelopes likely formed outside a few AU, since accreting such large gas mass fractions within typical disk lifetimes onto a water poor rocky cores is very challenging \citep{LF14,IS15,LC16}. 

With the launch of TESS and the James Webb Telescope around the corner, new discoveries await and further advances in field of super-Earth formation are within close reach.

\section{Cross-References}
\begin{itemize}
\item{Formation of Terrestrial Planets by Andre Izidoro \& Sean Raymond}
\item{Planetary migration in protoplanetary discs by Richard Nelson}
\item{Transit Timing and Duration Variations for the Discovery and Characterization of Exoplanets, by Eric Agol \& Daniel Fabrycky}
\item{Dynamical evolution of planetary systems by Alessandro Morbidelli}
\end{itemize}

\begin{acknowledgement}
The author thanks Sivan Ginzburg and John Biersteker for valuable comments and suggestions that contributed to this manuscript.
\end{acknowledgement}

\bibliographystyle{spbasicHBexo}

\end{document}